\documentclass[fp,twocolumn]{jpsj3}
\usepackage{txfonts}

\usepackage[dvipdfmx]{graphicx}
\usepackage{dcolumn}
\usepackage{bm}
\usepackage{verbatim}
\usepackage{wrapfig}
\usepackage{ascmac}
\usepackage{makeidx}
\usepackage{mathrsfs}
\usepackage{braket}
\usepackage{color}
\usepackage{ulem}

\newcommand{\nitio}{NiTiO$_{3}$}

\title{Dirac Magnon in Honeycomb Lattice Magnet \rm{NiTiO$_{3}$}}

\author{Hodaka Kikuchi$^1$\thanks{hodaka.kikuchi@issp.u-tokyo.ac.jp}, Makoto Ozeki$^1$, Nobuyuki Kurita$^2$, Shinichiro Asai,$^1$ Travis J. Williams$^3$, Tao Hong$^4$ and Takatsugu Masuda$^{1,5,6}$}
\inst{$^1$Institute for Solid State Physics, University of Tokyo, Kashiwa, Chiba 277-8581, Japan\\
$^2$Department of Physics, Institute of Science Tokyo, 2-12-1, Ookayama, Meguro-ku, Tokyo 152-8551, Japan\\
$^3$ISIS Pulsed Neutron and Muon Source, STFC Rutherford Appleton Laboratory, Harwell Campus, Didcot, OX11 0QX, United Kingdom\\
$^4$Neutron Scattering Science Division, Oak Ridge National Laboratory, Oak Ridge, Tennessee 37831-6393, USA\\
$^5$Institute of compounds Structure Science, High Energy Accelerator Research Organization, Ibaraki 305-0801, Japan\\
$^6$Trans-scale Quantum Science Institute, The University of Tokyo, Tokyo 113-0033, Japan} 

\abst{We performed inelastic neutron scattering experiments on single-crystal samples of the honeycomb lattice magnet, ilmenite NiTiO$_3$.
Below the N\'eel temperature of 22 K, spin wave excitations with a band energy of 3.7 meV were observed.
The neutron energy spectra were well-reproduced by modeling the system as a ferromagnetic honeycomb lattice with antiferromagnetic interlayer coupling,  using linear spin wave theory. 
Similar to another ilmenite CoTiO$_3$, a crossing structure was observed at the K point, suggesting the presence of Dirac magnons in NiTiO$_3$. 
Further calculations suggested the formation of Dirac nodal line.
}

\begin{document}
\maketitle

\section{Introduction}
The importance of topology in condensed matter physics has been widely recognized since the discovery of topological insulators. 
The compounds possess special metallic states at their edges or surfaces, where massless Dirac fermions give rise to the quantum Hall effect \cite{Kane2005,Bernevig2006,Markus2007,Hsieh2008}.
This discovery has opened up the potential for applications in highly efficient spintronics by utilizing spin currents generated at the edges or surfaces \cite{Moore2010}.
The concept of topology can be extended not only to fermion systems but also to magnon systems.
A representative example of this is the thermal Hall effect, where magnons carry heat flow perpendicular to a temperature gradient \cite{Katsura2010,Onose2010,Matsumoto2011}.
The Berry phase, induced by the Dzyaloshinskii-Moriya interaction, plays a crucial role in this phenomenon, and the imbalance of magnon edge currents leads to the Hall effect.

To date, the existence of topological magnons has been confirmed in various compounds through inelastic neutron scattering (INS) experiments.
In the kagome lattice Cu(1,3-bdc) \cite{Chisnell2015}, and in the layered honeycomb lattice compounds CrI$_3$ \cite{Chen2018}, CrBr$_3$ \cite{Cai2021}, and in the 3-dimensional (3D) centrosymmetric ferromagnet Mn$_5$Ge$_3$ \cite{Dias2023}, the bulk magnon dispersion at the K point is gapped, and theoretical calculations have demonstrated the existence of edge states inside the gap.
In Cu$_3$TeO$_6$ \cite{Yao2018,Bao2018}, both bulk and edge magnon dispersions cross at the K point, forming a Dirac cone, and the existence of Dirac magnons has been concluded. 
Similarly, in the ilmenite CoTiO$_3$ \cite{Yuan2020}, bulk and edge magnon dispersions cross, leading to the confirmation of Dirac magnons. 
The Dirac cone forms a Dirac nodal line (DNL) that continuously exists in momentum space.
As seen in fermion systems, topological structures are now being discovered one after another in magnon systems as well.

\begin{figure}[b]
\includegraphics[width=8.3cm]{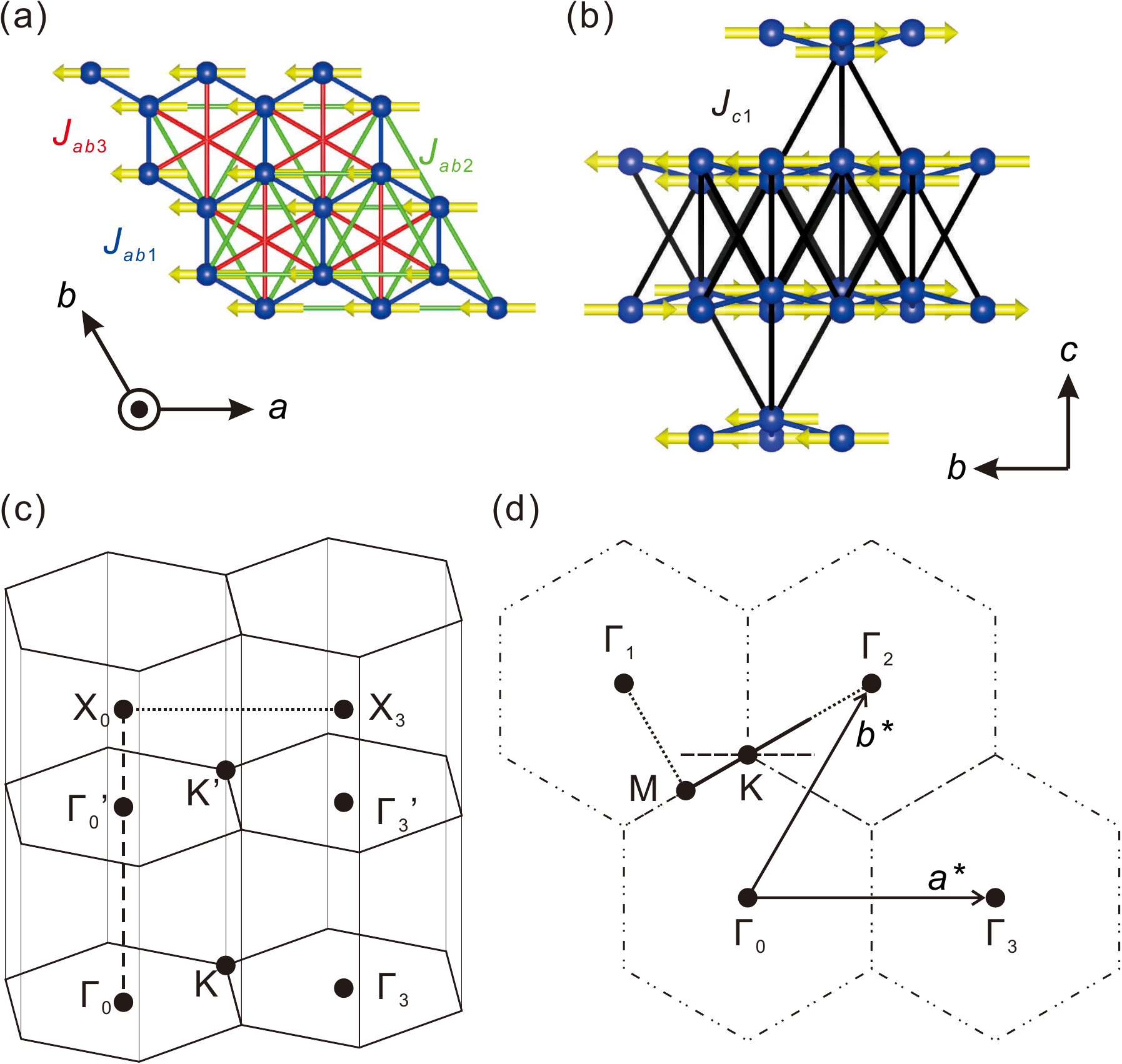}
\caption{
(a), (b) Crystal and magnetic structures of NiTiO$_3$. Ni$^{2+}$ atoms (3$d^{8}$, $S$ = 1) form a 
honeycomb lattice, with interaction paths indicated. The $a$-$b$ plane is illustrated in (a), while the $a$-$c$ plane is shown in (b).
(c) Schematic of the three-dimensional Brillouin zone (BZ). 
Inelastic neutron scattering (INS) spectra collected by the HODACA spectrometer were cut along the dotted and dashed lines.
(d) BZ in the two-dimensional honeycomb plane. 
INS spectra collected by the CTAX spectrometer were measured along the dotted, dashed, and solid lines.
}
\label{structure}
\end{figure}

The family of compounds $M$TiO$_3$ ($M$ = Mn, Fe, Co, Ni), which have an ilmenite structure, have long been studied \cite{Goodenough1967}. 
The crystal structure consists of alternating layers of TiO$_6$ and $M$O$_6$ octahedra along the $c$-axis, forming a structure similar to $ABC$ stacked graphene.
Thus far, bulk measurements \cite{Akimitsu1977,KATO1982,Hoffmann2021}, neutron diffraction \cite{Shirane1959}, and INS neutron scattering experiments \cite{Hwang2021,Kato1986,Yuan2020} have been conducted for $M$ = Mn, Fe, Co.
Additionally, in mixed crystals of MnTiO$_3$ and FeTiO$_3$, as well as NiTiO$_3$, phenomena such as spin glass have been observed \cite{Yoshizawa1989,Kawano1993}.
Recently, bulk measurements and neutron diffraction for NiTiO$_3$, where Ni$^{2+}$ ion has a spin $S = 1$, were reported \cite{Dey2020,Dey2021}. 
Below the N\'eel temperature ($T_{\rm N}$) of 22 K, NiTiO$_3$ exhibits magnetic order where the honeycomb lattice in the $a$-$b$ plane is ferromagnetic, and the layers along the $c$-axis are antiferromagnetically stacked, as shown in Figs.~\ref{structure}(a) and \ref{structure}(b)~\cite{Dey2021}.
This magnetic structure is the same as that of CoTiO$_3$. 
Furthermore, domains of ferroaxial order, which has recently attracted attention as a novel ordered state, have been observed \cite{Hayashida2020}. However, no studies on magnetic excitations have been reported in NiTiO$_3$. 
In this study, we performed INS experiments aimed at investigating magnetic excitations, searching for Dirac magnons, and identifying the spin Hamiltonian in NiTiO$_{3}$.

\section{Experimental details}

To collect neutron spectra in wide momentum ($q$) - energy ($\hbar\omega$) space, INS experiments were carried out by HOrizontally Defocusing Analyzer Concurrent data Acquisition (HODACA)~\cite{Kikuchi2024} installed at C11 beam port in JRR3. 
A single crystal sample of \nitio\ with a mass of 1.20 g was prepared. 
Gifford-McMahon (GM) cryostat was used to achieve the measurement temperature of 2.4 K.
The data reduction was performed by Advanced SYstem for User's data Reduction and Analysis (ASYURA) software. 

To detail the spectra in low-energy range, INS experiments were carried out by the cold neutron triple-axis spectrometer (CTAX) installed at High Flux Isotope Reactor (HFIR) in Oak Ridge National Laboratory (ORNL).
A single crystal with a mass of 1.50 g was prepared.
Low temperatures were achieved with a standard $^4$He cryostat, maintaining the measurement temperature at 1.4 K. 
The collimator setup was configured as guide - open - radial collimator - 120'.
Final neutron energy was fixed at 3.5 meV.
Be filter was employed downstream of the analyzer to cut off the higher harmonics.

\section{Experimental results}

\begin{figure}
\includegraphics[width=8.3cm]{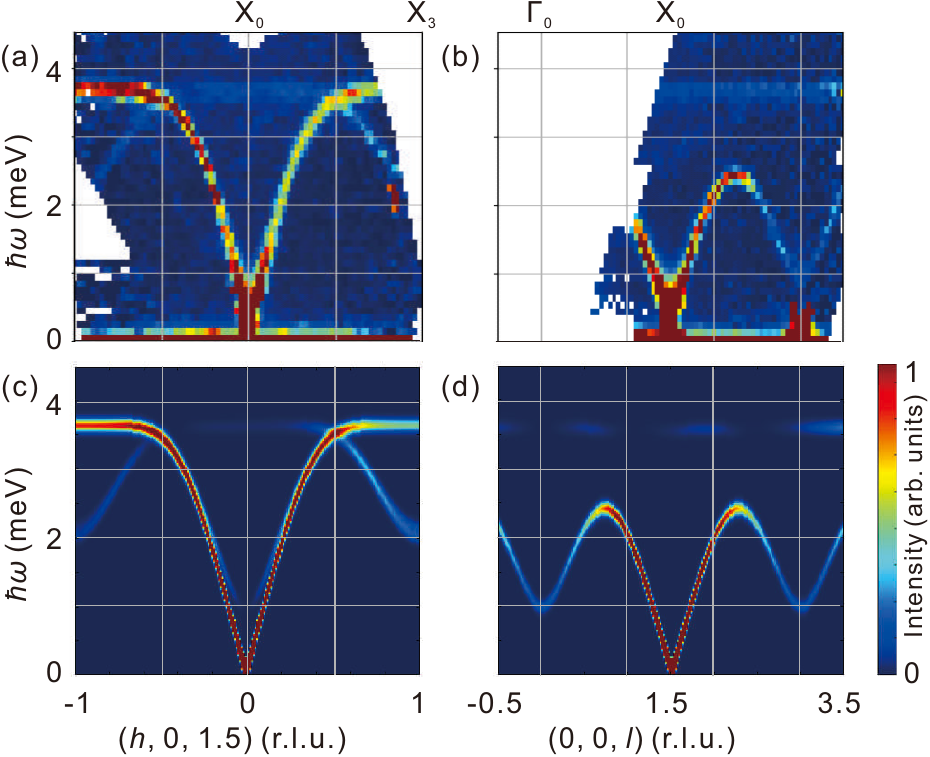}
\caption{
(a), (b) Inelastic neutron scattering spectra measured along the ($h$, 0, 1.5) direction in (a) and the (0, 0, $l$) direction in (b) at $T$ = 2.4 K, using the HODACA spectrometer.
(c), (d) Calculated inelastic neutron scattering spectra along the ($h$, 0, 1.5) direction in (c) and the (0, 0, $l$) direction in (d), based on linear spin-wave theory.
}
\label{HODACA_exp}
\end{figure}

The INS spectra in the $a^*$-$c^*$ plane were measured using the HODACA spectrometer. 
Figures \ref{HODACA_exp}(a) and \ref{HODACA_exp}(b) show the spectra cut along the lines in the 
reciprocal space, ($h$, 0, 1.5) and the (0, 0, $l$), respectively. 
The lines are shown by the dotted and dashed lines in Fig. \ref{structure}(c). 
The acoustic mode was observed rising from the magnetic propagation vector (${\bm q}_m$) of (0, 0, 1.5) up to 3.7 meV. 
In Fig. \ref{HODACA_exp}(a), the periodicity of the dispersion curve in the $a^*$ direction is longer than that of the reciprocal lattice unit. 
NiTiO$_3$ has a rhombohedral lattice with space group $R\bar{3}$, consisting of an $ABC$ stacked honeycomb lattice with centering by $[2/3,~1/3,~1/3]$ and $[1/3,~2/3,~2/3]$, leading to the dispersion curve following a 3$a^*$ periodicity.
In Fig. \ref{HODACA_exp}(b), an acoustic mode with a band energy of 2.5 meV was observed, which is about 2/3 of the band energy in the $a^*$ direction. 
The compound, thus, has strong interactions along the $c$ axis as well to be a 3D magnetic system. 
A flat optical mode was observed at 3.7 meV, consistent with the band energy in the $a^*$ direction.
The observed intensity near $\hbar\omega$ = 0 meV and at ${\bm q}$ = (0, 0, 3) is the nuclear 
Bragg peak intensity spreading into the energy direction because of the instrument resolution.
As for the inelastic channel, a gap excitation at $\hbar\omega$ = 1 meV was observed, suggesting the presence of easy-plane anisotropy.
The fact that the acoustic mode rises from (0, 0, 1.5) means that the compound has antiferromagnetic correlations along the $c$ axis, consistent with the reported magnetic structure\cite{Dey2021}.

\begin{figure}
\includegraphics[width=8.3cm]{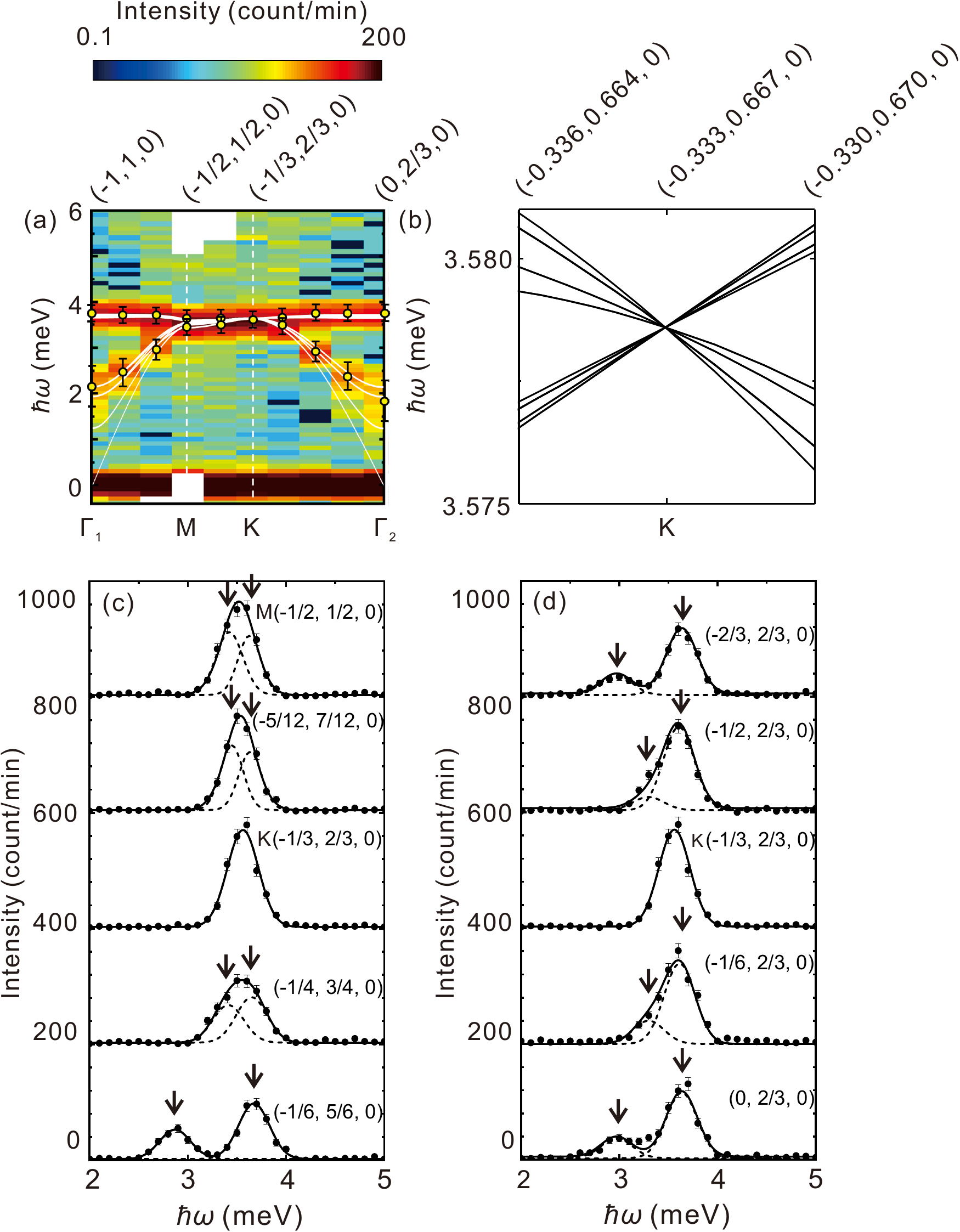}
\caption{
(a) Pseudo-color plot of the INS spectra measured along the dotted line in Fig. \ref{structure}(d) at $T = 1.4$ K using the CTAX spectrometer. The color bar is logarithmic. White lines represent dispersion curves calculated using linear spin-wave theory, and the dots indicate energy peak positions obtained by Gaussian fitting of the spectra. Error bars correspond to the full width at half maximum of the Gaussian fit.
(b) An enlarged view of the calculated dispersion curves around the K point.
(c), (d) INS spectra around the K point, measured along the solid line in Fig. \ref{structure}(d) in the (-1/3+$h$, 2/3+$h$, 0) direction for (c) and along the dashed line in the (-1/3, 2/3+$k$, 0) direction for (d). Solid lines represent the Gaussian fitting curves.
}
\label{CTAX_exp}
\end{figure}

The INS spectra in the $a^*$-$b^*$ plane were measured using the CTAX spectrometer. 
The spectra measured along the dotted line in reciprocal space in Fig. \ref{structure}(c) ($\Gamma_1$ - M - K - $\Gamma_2$), are presented in Fig. \ref{CTAX_exp}(a). 
The white lines represent the dispersion curve calculated using linear spin wave theory (LSWT). 
The circles represent the peak energies obtained through Gaussian fitting, and the error bars indicate the full width at half maximum (FWHM) of the peaks obtained from the fit.
The peaks observed at the $\Gamma$ point below 3 meV are broadened in the energy direction because of multiple nearby modes. 
Since the peak energies were too close to separate, a single Gaussian fit was applied. 
In the measured energy range, two peaks were observed. 
While they converge into a single peak at the K point, they do not converge at the M point.
Figures \ref{CTAX_exp}(c) and \ref{CTAX_exp}(d) investigate this crossing at the K point, showing the path along the BZ in Fig. \ref{structure}(d), respectively. 
At 3.6 meV at the K point, the modes clearly exhibit a crossing within the instrumental resolution.
The enlarged view of the calculated dispersion curves around K point in Fig.~\ref{CTAX_exp}(b) shows a linear crossing. 

Generally, in a honeycomb lattice with ferromagnetic ordering in the $ab$ plane, when the bulk dispersion curve shows a linear crossing, the surface state also exhibits a linear crossing, forming the so-called Dirac cone \cite{Pershoguba2018}. 
Therefore, the observed crossing of the dispersion curve in NiTiO$_3$ suggests the existence of a Dirac cone in the compound, suggesting the presence of Dirac magnons.

\section{Analysis}

We analyzed the INS spectra using LSWT. 
The spin Hamiltonian is given as follows;
\begin{eqnarray}
{\mathcal H}&=&J_{ab1}\sum_{i.j}\bm{S}_{i}\cdot\bm{S}_{j}+J_{ab2}\sum_{i,k}\bm{S}_{i}\cdot\bm{S}_{k}+J_{ab3}\sum_{i,l}\bm{S}_{i}\cdot\bm{S}_{l} \nonumber \\
&+&J_{c1}\sum_{i,m}\bm{S}_{i}\cdot\bm{S}_{m}+D\sum_{i}(\bm{S}_{z})^{2},
\label{eq1}
\end{eqnarray}
where the exchange interactions, $J_{ab1}$, $J_{ab2}$, $J_{ab3}$, and $J_{c1}$ are defined in Figs. \ref{structure}(a) and \ref{structure}(b). 
$D$ denotes a single ion anisotropy. 
The presence of the DM interaction is allowed  between next-nearest-neighbor Ni ions. It is, however, not considered in this analysis because the canting of the spins induced by DM interaction is not reported. 
The LSWT analysis was conducted using the spectrum measured in the $a^*$-$b^*$ plane with the CTAX spectrometer and in the $a^*$-$c^*$ plane with the HODACA spectrometer. 
The weighted residual sum of squares was calculated using the following equation;
\begin{eqnarray}
\chi^2=\frac{1}{N}\sum_{i,j}\frac{(\hbar\omega_{i}^{\rm{exp}}(\bm{q}_{j})-\hbar\omega_{i}^{\rm{cal}}(\bm{q}_{j}))^2}{\sigma_{i}^2}.
\label{eq2}
\end{eqnarray}
Here, $\hbar\omega_{i}^{\rm{exp}}(\bm{q}_{j})$ is the $i$th peak energy observed at $\bm{q} = \bm{q}_{j}$ via constant q-scan/slice, $\hbar\omega_{i}^{\rm{cal}}(\bm{q}_{j})$ is the calculated peak energy closest to $\hbar\omega_{i}^{\rm{exp}}(\bm{q}_{j})$, $\sigma_{i}$ is the FWHM obtained by Gaussian fitting of the experimental data, and $N$ represents the number of measured data points. 
The least squares method was used to optimize the five parameters in the spin Hamiltonian, within the –2 to 2 meV range.
Table \ref{tab1} shows the parameters that yield the smallest $\chi^{2}$. The parameter range where $\chi^{2}$ increases by a factor of 1.5 is defined as the error, shown in parentheses. 
The spectra calculated using these values are shown in Figs. \ref{HODACA_exp}(c) and \ref{HODACA_exp}(d), and the dispersion curves are shown in Fig. \ref{CTAX_exp}(a). 
Since the calculations reproduce the experimental results very well, the spin Hamiltonian is determined.

\begin{table}
\caption{
The parameters of the spin Hamiltonian. The unit is meV. 
}
\begin{tabular}{ccccc} \hline
$J_{ab1}$ & $J_{ab2}$ & $J_{ab3}$ & $J_{c1}$ & $D$\\
$-0.15(2)$ & $-0.05(1)$ & $-0.05(1)$ & $0.27(3)$ & $0.10(2)$\\ \hline
\end{tabular}
\label{tab1}
\end{table}

\begin{figure}
\includegraphics[width=8.3cm]{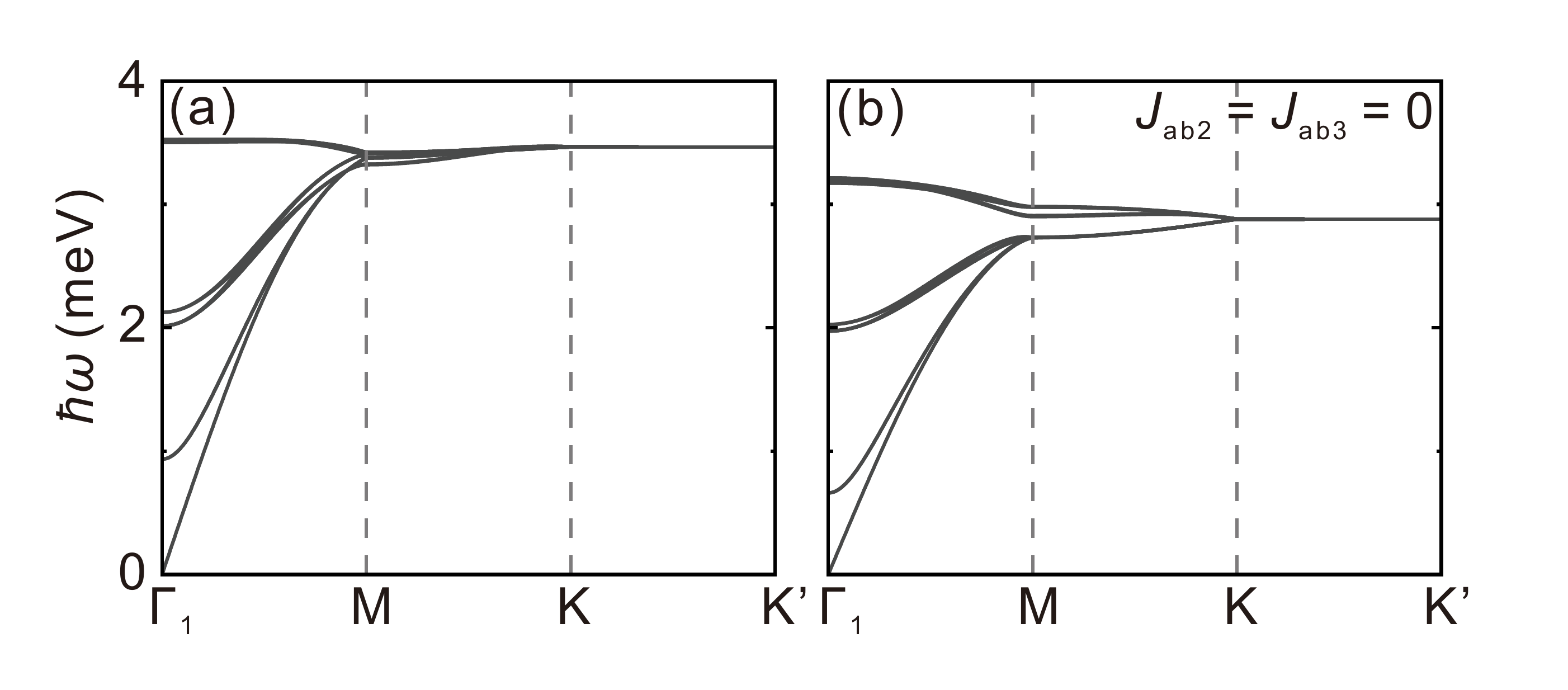}
\caption{
(a) Calculated dispersion curves using the parameters in Table \ref{tab1}. The modes converge at the K point, and a single flat dispersion exists between K and K'. (b) Calculated dispersion curves calculated with $J_{ab2} = J_{ab3} = 0$ while using other parameters from Table \ref{tab1}.}
\label{DNL_calc}
\end{figure}

\section{Discussion}

Figure \ref{DNL_calc}(a) shows the calculated dispersion curves using the parameters in Table \ref{tab1}. 
The spin wave modes approach each other from the $\Gamma$ point to the M point without intersecting; 
there is still a gap of about 0.05 meV at the M point. 
When moving from the M point to the K point, the modes become completely degenerate. 
Lastly, along the K-K' direction, there is no energy dependence of the modes regardless of large interlayer interaction $J_{c1}$. 
These features suggest that in NiTiO$_3$, as in the similar compound CoTiO$_3$~\cite{Yuan2020}, a DNL is continuously formed along the $c^*$ direction, with Dirac cones present.
Figure \ref{DNL_calc}(b) shows the calculated dispersion curves with $J_{ab2} = J_{ab3} = 0$ 
while using other parameters from Table \ref{tab1}. 
Dirac cone structure at K point is more clearly seen in the absence of the further neighbor interactions. 
The present study indicates that the DNL remains stable even when the interlayer interaction is large and second and third nearest-neighbor interactions within the $ab$ plane are introduced. 

\section{\label{sec:level1}Conclusion}

Inelastic neutron scattering experiments were performed on ilmenite NiTiO$_3$. 
Spin Hamiltonian was identified as ferromagnetic honeycomb lattice with antiferromagnetic interlayer coupling. 
A crossing was clearly observed at the K point, suggesting the presence of a Dirac magnon.  
Further calculation suggests the presence of a Dirac nodal line due to the flat dispersion along the $c^*$ direction at the K point.


\section*{Acknowledgment}

\begin{acknowledgment}
We are grateful to D. Kawana, T. Asami, and R. Sugiura for supporting us in the neutron scattering experiment at HODACA. 
The neutron experiment using HODACA at JRR-3 was carried out by the joint research in the Institute for Solid State Physics, the University of Tokyo (Proposal 23545).  
A portion of this research used resources at the High Flux Isotope Reactor, a DOE Office of Science User Facility operated by the Oak Ridge National Laboratory (ORNL).
Travel expenses for the neutron scattering experiments performed using CTAX at ORNL, USA was supported by the U.S.-Japan Cooperative Research Program on Neutron Scattering.
This project was supported by JSPS KAKENHI Grant No. 21H04441. 
\end{acknowledgment}


\begin{thebibliography}{10}

\bibitem{Kane2005}
C.~L. Kane and E.~J. Mele, Phys. Rev. Lett. {\bfseries 95},  146802 (2005).

\bibitem{Bernevig2006}
B.~A. Bernevig, T.~L. Hughes, and S.-C. Zhang, Science {\bfseries 314},  1757
  (2006).

\bibitem{Markus2007}
M.~K{\"o}nig, S.~Wiedmann, C.~Br{\"u}ne, A.~Roth, H.~Buhmann, L.~W. Molenkamp,
  X.-L. Qi, and S.~Zhang, Science {\bfseries 318},  766  (2007).

\bibitem{Hsieh2008}
D.~Hsieh, D.~Qian, L.~Wray, Y.~Xia, Y.~S. Hor, R.~J. Cava, and M.~Z. Hasan,
  Nature {\bfseries 452},  970 (2008).

\bibitem{Moore2010}
J.~E. Moore, Nature {\bfseries 464},  194 (2010).

\bibitem{Katsura2010}
H.~Katsura, N.~Nagaosa, and P.~A. Lee, Phys. Rev. Lett. {\bfseries 104},
  066403 (2010).

\bibitem{Onose2010}
Y.~Onose, T.~Ideue, H.~Katsura, Y.~Shiomi, N.~Nagaosa, and Y.~Tokura, Science
  {\bfseries 329},  297 (2010).

\bibitem{Matsumoto2011}
R.~Matsumoto and S.~Murakami, Phys. Rev. Lett. {\bfseries 106},  197202 (2011).

\bibitem{Chisnell2015}
R.~Chisnell, J.~S. Helton, D.~E. Freedman, D.~K. Singh, R.~I. Bewley, D.~G.
  Nocera, and Y.~S. Lee, Phys. Rev. Lett. {\bfseries 115},  147201 (2015).

\bibitem{Chen2018}
L.~Chen, J.-H. Chung, B.~Gao, T.~Chen, M.~B. Stone, A.~I. Kolesnikov, Q.~Huang,
  and P.~Dai, Phys. Rev. X {\bfseries 8},  041028 (2018).

\bibitem{Cai2021}
Z.~Cai, S.~Bao, Z.-L. Gu, Y.-P. Gao, Z.~Ma, Y.~Shangguan, W.~Si, Z.-Y. Dong,
  W.~Wang, Y.~Wu, D.~Lin, J.~Wang, K.~Ran, S.~Li, D.~Adroja, X.~Xi, S.-L. Yu,
  X.~Wu, J.-X. Li, and J.~Wen, Phys. Rev. B {\bfseries 104},  L020402 (2021).

\bibitem{Dias2023}
M.~d.~S. Dias, N.~Biniskos, F.~J. dos Santos, K.~Schmalzl, J.~Persson,
  F.~Bourdarot, N.~Marzari, S.~Bluegel, T.~Brueckel, and S.~Lounis, Nat.
  Commun. {\bfseries 14},  (2023).

\bibitem{Yao2018}
W.~Yao, C.~Li, L.~Wang, S.~Xue, Y.~Dan, K.~Iida, K.~Kamazawa, K.~Li, C.~Fang,
  and Y.~Li, Nat. Phys. {\bfseries 14},  1011^^e2^^80^^931015 (2018).

\bibitem{Bao2018}
S.~Bao, J.~Wang, W.~Wang, Z.~Cai, S.~Li, Z.~Ma, D.~Wang, K.~Ran, Z.-Y. Dong,
  D.~L. Abernathy, S.-L. Yu, X.~Wan, J.-X. Li, and J.~Wen, Nat. Commun.
  {\bfseries 9},  2591 (2018).

\bibitem{Yuan2020}
B.~Yuan, I.~Khait, G.-J. Shu, F.~C. Chou, M.~B. Stone, J.~P. Clancy,
  A.~Paramekanti, and Y.-J. Kim, Phys. Rev. X {\bfseries 10},  011062 (2020).

\bibitem{Goodenough1967}
J.~B. Goodenough and J.~J. Stickler, Phys. Rev. {\bfseries 164},  768 (1967).

\bibitem{Akimitsu1977}
J.~Akimitsu and Y.~Ishikawa, J. Phys. Soc. Jpn. {\bfseries 42},  462 (1977).

\bibitem{KATO1982}
H.~Kato, M.~Yamada, H.~Yamauchi, H.~Hiroyoshi, H.~Takei, and H.~Watanabe, J.
  Phys. Soc. Jpn. {\bfseries 51},  1769 (1982).

\bibitem{Hoffmann2021}
M.~Hoffmann, K.~Dey, J.~Werner, R.~Bag, J.~Kaiser, H.~Wadepohl, Y.~Skourski,
  M.~Abdel-Hafiez, S.~Singh, and R.~Klingeler, Phys. Rev. B {\bfseries 104},
  014429 (2021).

\bibitem{Shirane1959}
G.~Shirane, S.~J.~Pickart, and Y.~Ishikawa, J. Phys. Soc. Jpn. {\bfseries 14},
  1352 (1959).

\bibitem{Hwang2021}
I.~Y. Hwang, K.~H. Lee, J.-H. Chung, K.~Ikeuchi, V.~O. Garlea, H.~Yamauchi,
  M.~Akatsu, and S.~Shamoto, J. Phys. Soc. Jpn. {\bfseries 90},  064708
  (2021).

\bibitem{Kato1986}
H.~Kato, Y.~Yamaguchi, M.~Yamada, S.~Funahashi, Y.~Nakagawa, and H.~Takei,
  Journal of Physics C: Solid State Physics {\bfseries 19},  6993 (1986).

\bibitem{Yoshizawa1989}
H.~Yoshizawa, S.~Mitsuda, H.~Aruga, and A.~Ito, J. Phys. Soc. Jpn. {\bfseries
  58},  1416 (1989).

\bibitem{Kawano1993}
H.~Kawano, H.~Yoshizawa, A.~Ito, and K.~Motoya, J. Phys. Soc. Jpn. {\bfseries
  62},  2575 (1993).

\bibitem{Dey2020}
K.~Dey, S.~Sauerland, J.~Werner, Y.~Skourski, M.~Abdel-Hafiez, R.~Bag,
  S.~Singh, and R.~Klingeler, Phys. Rev. B {\bfseries 101},  195122 (2020).

\bibitem{Dey2021}
K.~Dey, S.~Sauerland, B.~Ouladdiaf, K.~Beauvois, H.~Wadepohl, and R.~Klingeler,
  Phys. Rev. B {\bfseries 103},  134438 (2021).

\bibitem{Hayashida2020}
T.~Hayashida, Y.~Uemura, K.~Kimura, D.~Matsuoka, S. and~Morikawa, S.~Hirose,
  K.~Tsuda, T.~Hasegawa, and T.~Kimura, Nat. Commun. {\bfseries 11},  (2020).

\bibitem{Kikuchi2024}
H.~Kikuchi, S.~Asai, T.~J. Sato, T.~Nakajima, L.~Harriger, I.~Zaliznyak, and
  T.~Masuda, J. Phys. Soc. Jpn. {\bfseries 93},  091004 (2024).

\bibitem{Pershoguba2018}
S.~S. Pershoguba, S.~Banerjee, J.~C. Lashley, J.~Park, H.~\AA{}gren, G.~Aeppli,
  and A.~V. Balatsky, Phys. Rev. X {\bfseries 8},  011010 (2018).

\end{thebibliography}

\end{document}